%% file: main.tex
\documentclass[11pt]{article}

\usepackage[margin=1in]{geometry}

\usepackage{amsmath,amssymb,amsfonts}
\usepackage{graphicx}
\usepackage[hidelinks]{hyperref}
\usepackage{multirow}
\usepackage{booktabs}
\usepackage{tabularx}
\usepackage[ruled,vlined]{algorithm2e}
\usepackage{xcolor}
\usepackage{mdframed}
\usepackage{adjustbox}
\usepackage{url}
\usepackage{float}
\usepackage{placeins}
\usepackage{microtype}
\usepackage{ragged2e}
\justifying

\graphicspath{{figures/}}

{%
  \begin{mdframed}[backgroundcolor=gray!10, linecolor=white, linewidth=0pt]
  \bfseries TAKEAWAY.~\normalfont
}%
{\end{mdframed}}

\newenvironment{takeawaybox}%
{%
  \begin{mdframed}[backgroundcolor=blue!10, linecolor=white, linewidth=0pt,
    innerleftmargin=6pt, innerrightmargin=6pt,
    innertopmargin=6pt, innerbottommargin=6pt]
  \textbf{NOTE.}\ \normalfont
}%
{\end{mdframed}}

\definecolor{olivegreen}{rgb}{0.6, 0.73, 0.35}

\newenvironment{limitationbox}%
{%
  \begin{mdframed}[backgroundcolor=olivegreen!20, linecolor=white, linewidth=0pt,
    innerleftmargin=6pt, innerrightmargin=6pt,
    innertopmargin=6pt, innerbottommargin=6pt]
  \textbf{LIMITATIONS.}\ \normalfont
}%
{\end{mdframed}}

\title{Behavior-Aware and Generalizable Defense Against Black-Box Adversarial Attacks for ML-Based IDS}

\author{
Sabrine Ennaji \\
\small Sapienza University of Rome, Italy \\
\small \texttt{ennaji@di.uniroma1.it}
\and
Elhadj Benkhelifa \\
\small Staffordshire University, United Kingdom \\
\small \texttt{e.benkhelifa@staffs.ac.uk}
\and
Luigi Vincenzo Mancini \\
\small Sapienza University of Rome, Italy \\
\small \texttt{mancini@di.uniroma1.it}
}

\date{}

\begin{document}
\maketitle

\begin{abstract}
\input{abstract}
\end{abstract}

\vspace{0.5em}

\input{introduction}

\input{related_work}
\input{threat_model}

\input{defense}
\input{discussion}
\input{conclusion}

\bibliographystyle{unsrt}
\bibliography{bibliography}

\end{document}

%% file: abstract.tex
\begin{abstract}
Machine learning-based Intrusion Detection Systems (IDS) are increasingly targeted by black-box adversarial attacks, where attackers craft evasive inputs using indirect feedback such as binary outputs or behavioral signals like response time and resource usage. While several defenses have been proposed; including input transformation, adversarial training, and surrogate detection, they often fall short in practice. Most are tailored to specific attack types, require internal model access, or rely on static mechanisms that fail to generalize across evolving attack strategies. Furthermore, defenses like input transformation can degrade IDS performance, making them unsuitable for real-time deployment.

To address these limitations, we propose Adaptive Feature Poisoning (AFP), a lightweight and proactive defense mechanism designed specifically for realistic black-box scenarios. AFP assumes that probing can occur silently and continuously, and introduces dynamic, context-aware perturbations to selected traffic features corrupting the attacker’s feedback loop without impacting the IDS’s detection capabilities. AFP leverages traffic profiling, change point detection, and adaptive scaling to selectively perturb features the attacker is likely exploiting, based on observed deviations.

We evaluate AFP against diverse methods of realistic adversarial attacks including silent probing attacks, transferability- and decision boundary-based attacks, demonstrating its ability to confuse the attacker, degrade attack effectiveness, and preserve IDS performance. By offering a generalizable, attack-agnostic, and undetectable defense, AFP represents a significant step toward practical and robust adversarial resilience in real-world network environments.

\end{abstract}

%% file: introduction.tex
\section{Introduction}

Modern cybersecurity infrastructures now rely heavily on Intrusion Detection Systems (IDS) that are based on machine learning (ML) \cite{berguiga2025hids,wu2024intelligent}. By monitoring massive amounts of network traffic to identify patterns and abnormalities that might indicate attacks, these systems provide dynamic, intelligent protection capabilities. Unlike traditional rule-based IDS, which fail to deal with the emerging strategies of cyber adversaries, ML-based IDS adapt to new attack techniques and can detect previously unseen behaviors, making them highly valuable in today’s complex and high-throughput network environments \cite{vinayakumar2019deep}.

Despite these advantages, ML-based IDS are increasingly susceptible to adversarial attacks—carefully created inputs aimed at deceiving the model \cite{he2023adversarial}. Many defenses against such attacks have been provided by the academic community \cite{vitorino2023sok, pawlicki2020defending}. However, most of these solutions are developed and evaluated under simplified, controlled conditions, usually assuming full white-box access, where the attacker has knowledge of the model's internal workings \cite{apruzzese2022modeling}. Therefore, this lacks realism in modeling practical attack behavior. In real world settings, adversaries frequently operate in black-box environments using side channels like system response time, packet delay, or resource consumption to infer system performance. The effectiveness and reliability of current defenses are significantly compromised by this disconnect between theoretical assumptions and real-life conditions. There is a broad agreement on the limitations of academic adversarial defenses, according to a new study conducted by Microsoft researchers that involved interviews with 28 companies in a variety of industries, including cybersecurity, healthcare, government, banking and agriculture \cite{kumar2020adversarial}. Organizations expressed serious concerns about the relevance of existing research and emphasised the urgent need to expand beyond traditional attack models to effectively address adversarial manipulation challenges in the real world.

Motivated by these reality-based constraints and industry feedback, we propose a new defense approach designed for practical deployment: Adaptive Feature Poisoning (AFP). The risk of silent probing attacks, in which adversaries observe system activity passively without triggering alerts, has been specifically addressed by AFP. These attacks employ side-channel indicators to determine the behavior of an intrusion detection system and to create disruptions that can evade detection. AFP disrupts this process by operating silently in the background, introducing minor, targeted perturbations to sensitive traffic features. These deceptions corrupt the information the attacker gathers, making it unreliable and preventing them from effectively modeling or bypassing the IDS. In contrast to current defenses that rely on input modifications, retraining, or anomaly detection modules, AFP is adaptive, lightweight, and developed for real-time deployment.  Without degrading the IDS's capacity to identify legitimate traffic, it actively influences the attacker's learning process.

This paper's main contributions are:

\begin{itemize}

\item We identify and formalize a silent probing threat in black-box adversarial environments, where attackers rely on indirect observations instead of direct interactions with the model.

\item We propose Adaptive Feature Poisoning (AFP); a real-time, lightweight defense strategy that dynamically injects adaptive perturbations into sensitive traffic features in response to side-channel activity. AFP uses change-point detection and signal monitoring to continuously disrupt the attacker’s ability to model the IDS behavior.

\item We evaluate AFP against different black-box adversarial attacks, such as probing techniques, transferability and decision boundary-based attacks. Our findings show that AFP effectively prevents successful evasion without sacrificing the accuracy of IDS.

\end{itemize}

The rest of this paper is organized as follows: 
Section \ref{sec:review} reviews the landscape of adversarial threats and existing defense strategies in ML-based IDS. Section \ref{sec:threat_model} defines the threat model and outlines the attacker’s capabilities. Section \ref{sec:defense} presents our proposed defense mechanism, (AFP), detailing its architecture, design rationale, and implementation algorithms for reproducibility. Section \ref{sec:discussion} presents the experimental setup and reports the evaluation results, and provides an in-depth analysis of AFP’s effectiveness, including a comparison with existing defense strategies. Finally, Section \ref{sec:conclusions} concludes the paper and discusses directions for future research.

%% file: related_work.tex
\section{Adversarial Threat Landscape and Defense Limitations in ML-Based IDS} \label{sec:review}
With a particular focus on black-box attack techniques, this section examines the evolving landscape of adversarial attacks targeting ML-based IDS. It proceeds by outlining the fundamental ideas of adversarial attacks in IDS contexts and explaining the relevance of black-box settings in practical implementations. Then, the main categories of existing defense strategies in the literature are critically analyzed.

\subsection{From Image-based Adversarial Attacks to Structured Network Traffic}

The field of image classification was the first to investigate adversarial attacks in machine learning, as subtle, undetectable modifications to input images might significantly change a model's predictions \cite{szegedy2013intriguing}.  An adversarial example for a classifier $f: \mathbb{R}^d \to \{1, \dots, K\}$ mapping an input $x \in \mathbb{R}^d$ (e.g, image) to a class label aims to find a perturbed input $x' = x + \delta$ such that:

\[
f(x') \neq f(x) \quad \text{with} \quad \|\delta\|_p \leq \epsilon
\]

where $\delta$ is limited to stay inside a small $l_p$-norm ball, ensuring that the perturbation remains human-imperceptible.  Such weaknesses exposed the vulnerability of even the most precise deep neural networks and spurred a surge of research on strong defenses and modeling \cite{goodfellow2014explaining, madry2018towards}.

Over time, the same principle was extended to network security, particularly in ML-based IDS, where models classify network traffic as malicious or benign \cite{javaid2016deep, yang2020network}. However, this transition from image-based \textit{(unstructured)} data to network-based \textit{(structured)} data presented significant challenges. 

In contrast to images, which are composed of uniform pixel grids, network traffic consists of structured features that are heterogeneous and frequently combine continuous and categorical attributes.   Formally, a network traffic sample can be described as:

\[
x \in \mathcal{X}_{\text{continuous}} \times \mathcal{X}_{\text{categorical}}
\]

where $\mathcal{X}_{\text{continuous}}$ includes numerical features (e.g., packet size, duration) and $\mathcal{X}_{\text{categorical}}$ contains discrete fields (e.g., protocol types, service ports).

Adversarial perturbations in this setting must maintain protocol specifications and functional integrity by satisfying small-magnitude limitations, as well as semantic and operational validity requirements. Violations may cause communication problems or immediately cause anomaly detection.  As a result, network security becomes much more limited and complex than typical unstructured domains when it comes to developing effective adversarial attacks and associated defensive measures \cite{apruzzese2022modeling}.



\begin{figure}[htbp]
  \centering
  \includegraphics[trim=2cm 2cm 2cm 1.75cm, clip, width=0.3\textwidth]{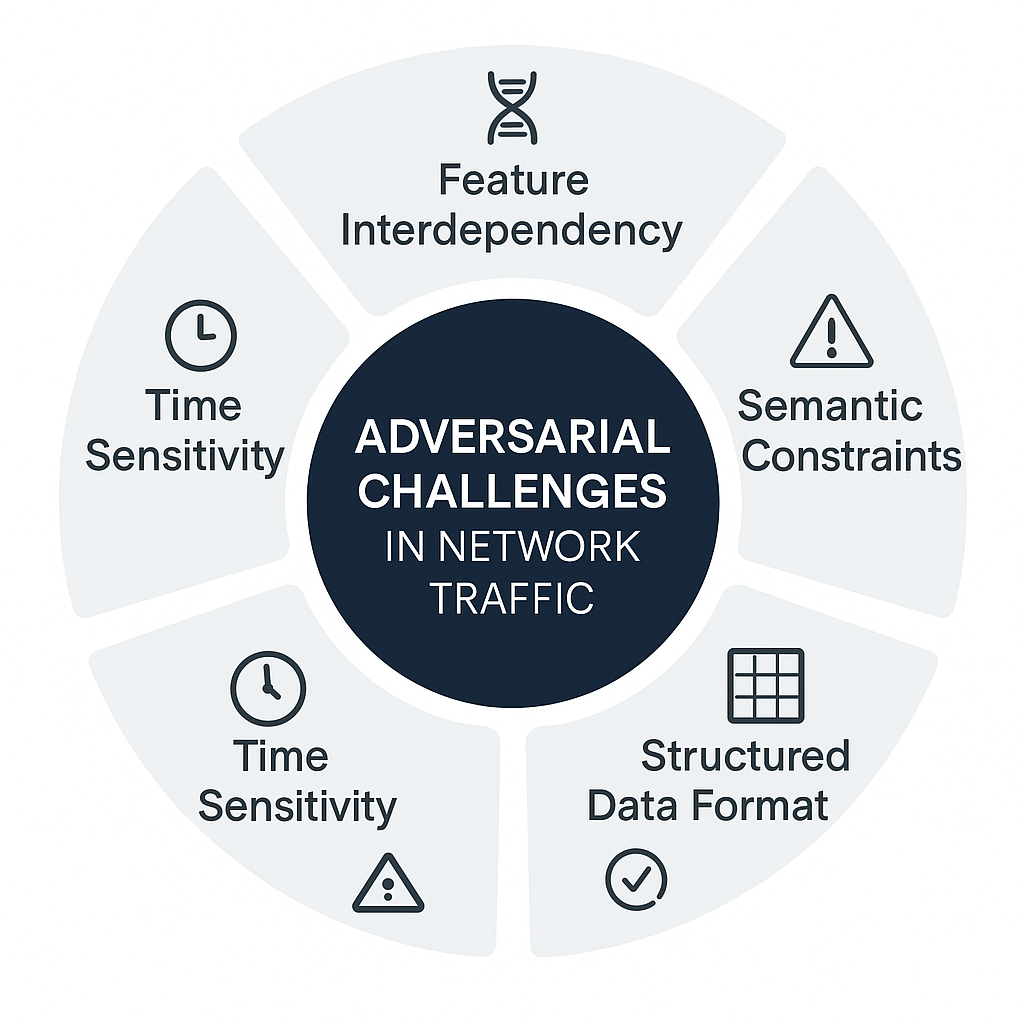}
  \caption{Adversarial Challenges in Network Traffic}
  \label{fig:adv_challenges}
\end{figure}

\subsubsection{Challenges of Adversarial Attacks in Network Security}
Generating adversarial techniques in the context of network traffic introduces different challenges that do not exist in unstructured data \textit{(e.g., image classification)}:

\begin{itemize}
\item \textbf{Feature interdependency:} Features of network traffic, such as time, bytes, and packet count, can frequently be causally and temporally interdependent. Perturbing one feature may impact others or trigger detection flags.
\item  \textbf{Semantic constraints:} Unlike image pixels, network traffic features frequently encode protocol-specific semantics (e.g., TCP flags, port numbers, sequence lengths). Randomly modifying such features may interrupt communication flows, violate protocol rules, or cause anomalies that protocol analyzers can identify. Therefore, adversarial manipulations must be generated carefully to ensure they maintain functional validity while remaining undetectable.
\item \textbf{Structured data format:} Inputs are represented as structured tabular records, often containing a combination of continuous and categorical values. Hence, this makes the perturbation less flexible. 
\item  \textbf{Time sensitivity:} Real-time traffic analysis is common. To avoid detection, adversarial manipulation-induced delays must be undetectable.
\item \textbf{Limited observability \& detection risk :} Adversaries cannot access internal workings of the target model in black-box scenarios. Thus, attack crafting is more challenging and dependent on indirect cues since they have to infer system behavior from limited feedback (usually binary labels or side-channel indicators). However, other methods like frequent probing used to extract this feedback can easily trigger alerts, making efficient and invisible adversarial perturbations  particularly challenging.
\end{itemize}

\begin{takeawaybox}These challenges call for new types of adversarial strategies and more robust, real-time defenses \cite{ennaji2025adversarial}.\end{takeawaybox}

\subsubsection{Multi-Dimensional Taxonomy of Adversarial Attacks}

\begin{figure*}[htbp]
  \centering
  \includegraphics[trim=0.3cm 13cm 0cm 11cm, clip, width=0.75\textwidth]{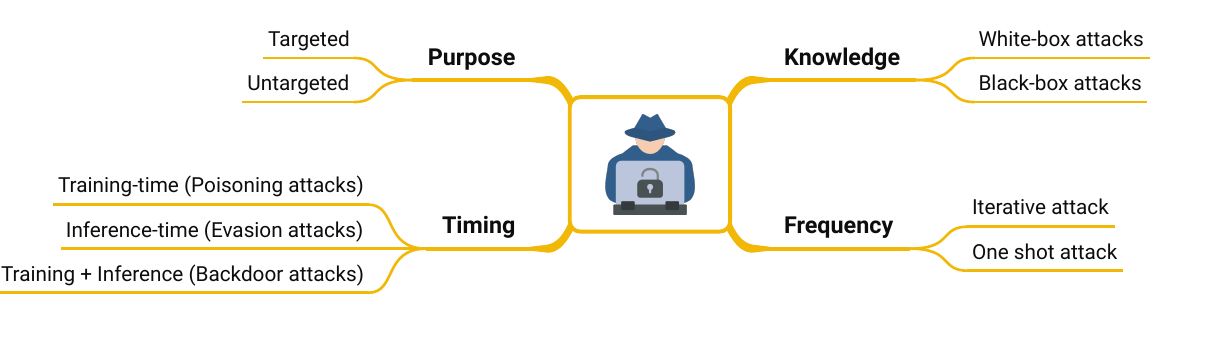}
  \caption{Multi-dimensional taxonomy of adversarial attacks against machine learning models organized by by attacker knowledge, purpose, timing of the attack and its frequency.}
  \label{fig:taxonomy}
\end{figure*}

Adversarial attacks against machine learning-based models can be classified based on a common multidimensional taxonomy, as summarized in Figure \ref{fig:taxonomy}.

\begin{itemize}
\item \textbf{By attacker knowledge:} The degree of access to the details of the target model defines the classification of the adversary's capabilities.
\begin{itemize}
    \item \textit{White-box attacks:} The architecture, parameters, and training data of the model are all fully visible to the adversary. Although frequently found in academic literature, these assumptions are rarely valid in realistic IDS deployments. There are numerous techniques for creating white-box adversarial attacks, including the Jacobian-based Saliency Map Attack \cite{linardatos2020explainable}, Deepfool \cite{moosavi2016deepfool}, Carlini \& Wagner \cite{carlini2017towards}, Projected Gradient Descent \cite{kurakin2018adversarial}, Fast Gradient Sign Method \cite{goodfellow2014explaining}, etc.
    \item \textit{Black-box attacks:} The attacker lacks access to any internal model features (architecture, parameters, and training data), which better reflect real-world deployment conditions. These attacks frequently use methods such as Zeroth-Order Optimization attacks \cite{chen2017zoo} or Generative Adversarial Networks (GAN) \cite{martins2020adversarial}. However, they can also be used on surrogate models by simulating the behavior of the target model and then using the learned models to generate adversarial inputs for the original system \cite{zhou2020dast}.
\end{itemize}

\item \textbf{By attack timing:} Attacks are classified according to when they occur.
\begin{itemize}
    \item \textit{Poisoning attacks:} These attacks are introduced during the training phase of a model and aim to alter learned patterns or decision boundaries by injecting modified samples into the training dataset \cite{tian2022comprehensive}. 
    \item \textit{Evasion attacks:} These are executed during a test time and involve subtle modifications to the input that are meant to avoid detection while maintaining the intended functionality of the sample \cite{yang2021rumor, martins2020adversarial}.
    \item \textit{Backdoor attacks:} These combine evasion and poisoning techniques throughout the training and inference phases \cite{liu2018trojaning}. The adversary injects a hidden trigger during training, which remains inactive until a specific input pattern activates it during inference; such as a crafted combination of protocol fields and packet sizes in network traffic that results in misclassification.
\begin{takeawaybox} Backdoor and poisoning attacks are continuing risks, even if they are less frequent in deployed setups and require access to the training pipeline. Evasion attacks, on the other hand, are the most thoroughly researched in academic literature because they are test-time focused and realistic, matching the capabilities of real-world attackers.
\end{takeawaybox}
\end{itemize}

\item \textbf{By purpose:} Adversarial perturbations can be either targeted or untargeted. 
\begin{itemize}
  \item \textit{Targeted attacks:} designed to incorrectly classify inputs into specific categories.
  \item \textit{Untargeted attacks:} Aimed at causing general misclassification without targeting a specific label.
\end{itemize}

\item \textbf{Attack frequency:} Iterative or one-shot adversarial attacks are both possible.  One-shot attacks use a single-step technique to create adversarial perturbations.  Multiple iterations are used in iterative adversarial attacks, which enhance the adversarial perturbation by continually querying the classifier.

\end{itemize}
While this taxonomy is also applicable in other domains (image, text, etc.), the operational environment of network security presents important differences in risk and feasibility. Despite widely studied in academic research, white-box attacks are unrealistic in practice due to the absence of internal model access in deployed IDS settings. In contrast, black-box evasion techniques reflect real-world threats since they align more with adversarial capabilities, often exploiting limited or indirect feedback in a stealthy way.

\subsubsection{Focus on Black-Box Attack Strategies}

Although black-box attacks have not received as much attention in the literature compared to white-box attacks, they represent a more realistic threat model for deployed IDS systems. The majority of current research assesses their defenses in hypothetical scenarios, which are rarely applicable in real-world settings and assume complete access to model parameters, gradients, or training data. Therefore, the proposed defenses often fail when applied in real-world settings, where attackers operate under more constrained, black-box conditions.

A variety of black-box attack techniques have been discovered in the literature and can be categorized as follows.

\begin{itemize}
\item \textbf{Query-based attacks:} These involve generating adversarial examples based on the observed responses after directly querying the target model to gather feedback. In real-world systems, they are moderately feasible, but in order to evade detection by IDS devices, they must achieve a balance between stealth and query performance \cite{bai2023query}.
\item \textbf{Decision boundary-based attacks:} These methods estimate the model's decision boundaries using optimization techniques and examine the model's output for carefully selected inputs. Adversarial perturbations are generated to subtly breach these boundaries without triggering detection \cite{finlay2019logbarrier}. While they often require a large number of adaptive queries to operate quietly, they are highly practical in black-box environments where only final decisions are revealed.
\item \textbf{Transferability-based attacks:} These involve generating adversarial examples on a surrogate model that has been trained on a similar dataset in the expectation that they will transfer to the target model \cite{wang2023towards, zhang2025explainable}. While widely used in research, this approach frequently fails in complicated domains such as network traffic, where variations in feature distributions and dependencies limit the success of transferability.
\item \textbf{Randomized \& gradient free-based attacks:} These techniques perform well against hardened models with obfuscated gradients because they rely on random perturbations or black-box optimization algorithms that do not require gradient knowledge, which makes them appropriate for situations when model gradients are inaccessible \cite{xiang2023improving}. Although feasible, they usually require many queries and take time to converge, which raises the risk of detection.
\end{itemize}

System monitoring, structured input dependencies, and detection risks limit the effectiveness of various black-box adversarial strategies in practical network traffic scenarios \cite{ennaji2025adversarial}. However, even a small amount of success in obtaining input or examining decision limits could seriously compromise IDS security.  This motivates the development of the Adaptive Feature Poisoning (AFP) framework, which is a proactive and adaptive defense that attempts to deceive attackers early in the reconnaissance phase without compromising IDS performance.

\subsection{Defense Strategies and Limitations in Adversarial IDS Research}
In this section, existing defense techniques for adversarial attacks in ML-based IDS, are categorized into two main groups: (i) Training-based defenses and (ii) Structure-based defenses. We analyze representative approaches in each group, identifying their assumptions, limitations, and gaps in adaptability, generalizability, and operational feasibility, as summarized in Table \ref{tab:defense_summary}.

This analysis lays the groundwork for the motivation behind our proposed defense strategy, Adaptive Feature Poisoning (AFP), which directly addresses the limitations of existing solutions.

\renewcommand{\arraystretch}{1.5}
\begin{table*}[h!]
\centering
\caption{Summary of Adversarial Defense Strategies and Their Limitations}
\label{tab:defense_summary}

\resizebox{\textwidth}{!}{%
\begin{tabular}{p{1.4cm}|p{4cm}|p{13cm}}
\hline
\textbf{References} & \textbf{Defense Categories} & \textbf{Limitations} \\
\hline
\cite{alahmed2022mitigation} & \multirow{4}{*}{\textbf{\textit{Adversarial Training}}} & Limited adaptability and generalizability due to reliance on a fixed GAN and narrow evaluation scope, with high computational cost. \\
\cline{1-1} \cline{3-3}
\cite{debicha2023adv} &  & Effectiveness depends on known attack patterns, assumes strong surrogate-target similarity, and incurs high retraining overhead. \\
\cline{1-1} \cline{3-3}
\cite{li2023eifdaa} &  & Relies on static adversarial patterns, small dataset limits generalization, and retraining costs reduce real-time feasibility. \\
\cline{1-1} \cline{3-3}
\cite{wu2025boosting} &  & Complexity of dynamic memory and adversarial sample generation limits real-time adaptability and operational feasibility. \\
\hline
\cite{zhang2020tiki} & \multirow{3}{*}{\textbf{\textit{Ensemble Learning}}} & Limited to decision-based attacks and not adaptive to evolving threats or diverse traffic. \\
\cline{1-1} \cline{3-3}
\cite{mccarthy2023defending} &  & Hierarchical structure's effectiveness depends on initial classification accuracy and sufficient per-class data. \\
\cline{1-1} \cline{3-3}
\cite{jiang2022fgmd} &  & Evaluation is constrained to IoT-specific scenarios and lack of generalizability. \\
\hline
\cite{roshan2024boosting} & \textbf{\textit{Input Transformation}} & Shows inconsistent results across datasets and classifiers, with potential trade-offs between robustness and accuracy. \\
\hline
\cite{hu2025tnd} & \multirow{3}{*}{\textbf{\textit{Detection and Rejection}}} & Manual tuning, false positives, and limited generalization hinder adaptability and scalability. \\
\cline{1-1} \cline{3-3}
\cite{de2020intrusion} &  & Focused on evasion attacks without adaptability to other types or changing hacker behaviors. \\
\cline{1-1} \cline{3-3}
\cite{pawlicki2020defending} &  & Static rules and lack of adaptive learning limit robustness in real-world, dynamic settings. \\
\hline
\cite{jmila2022adversarial} & \multirow{2}{*}{\textbf{\textit{Feature-space Regularization}}} & Performance varies across datasets; robustness gains often reduce accuracy. \\
\cline{1-1} \cline{3-3}
\cite{chen2023feddef} &  & Trade-off between privacy and performance impacts utility in real-time federated NIDS. \\
\hline
\cite{wang2022manda} & \textbf{\textit{Manifold Projection}} & Limited to specific attacks and datasets, with unclear adaptability to complex or unseen threats. \\
\hline
\end{tabular}
} 
\end{table*}

\begin{enumerate}
\item \textbf{Training-based defenses:} 

Training-based defenses improve the resilience of the model by explicitly integrating adversarial examples into the learning process. These strategies aim to expose the model to attack patterns during training phase, enabling enhanced generalization in adversarial scenarios. Typically, the training data is combined with adversarial examples that are created using well-known attack strategies, including gradient-free optimization, PGD, or FGSM. Although this method can increase the model’s ability to resist specific perturbations, its effectiveness is tightly related to the diversity, quality, and realism of the adversarial inputs used during training process. A representative example, by Alarmed et al. \cite{alahmed2022mitigation}, within this category employs Generative Adversarial Networks (GANs) to create adversarial traffic samples from the CSE-CICIDS2018 dataset, using 2000 training epochs. Using PCA for feature reduction, the model is trained in two phases: initially on clean data, and then on a combination of clean and adversarial samples generated by GAN. Despite the fact that the method is more resilient to the ZOO black-box attack, it has limited adaptability (due to reliance on a single GAN generator), weak generalizability because the evaluation scope is limited to a single attack strategy and model setup, and poor operational feasibility given its high computational requirements. 

Moreover, Debicha et al. \cite{debicha2023adv} combine adversarial training with a modular detection technique to protect ML-based NIDS against evasion attacks. It detects malicious inputs by comparing the outputs of a baseline model (trained on clean data) with a robust model (trained on adversarial examples) using an MLP-based IDS. Integration with different detection models is supported by ensemble techniques such as bagging. Despite its effictiveness, the framework's adaptability is restricted by its reliance on well-known attack patterns, and its generalizability weakens when used in various traffic scenarios. Concerns about operational feasibility are also raised by the increased complexity, which necessitates regular updates and incurring overhead. Additionally, the evaluation makes the assumption that the black-box threat model is transferability-based and depends on surrogate-target similarity, which may not be valid in real-world implementations.

Another adversarial training approach namely EEIFDAA \cite{li2023eifdaa} is proposed to strengthen IDSs in Industrial IoT (IIoT) settings by initially testing current IDS against adversarial methods, then retraining them using a combination of adversarial and original samples. Additionally, the system simulates evasive threats that lack typical malicious characteristics based on function-discarding adversarial samples. The defense has significant limitations even though it increases detection accuracy in controlled environments. Reliance on static adversarial patterns reduces adaptability and may fail against evolving attack strategies. Due to evaluation on a small dataset (e.g., X-IIoTID), generalizability is limited, which decreases effectiveness in various IIoT scenarios. The requirement for continuous retraining and the high computing expenses also affect operational feasibility. Despite the threat model that reflect realistic IIoT conditions, oversimplifying adversarial behavior through function-discarding may decreases realism and limit long-term robustness.

Recently, Adversarial training is repurposed by AdvAs-IIDS \cite{wu2025boosting} to improve learning in incremental intrusion detection, rather than just as a defensive method.  In order to enable the model maintain recognition performance on old classes while adjusting to new ones, it provides adversarial samples using a decoupled network architecture that maintains class-specific features. While a weighted cross-entropy loss reduces catastrophic forgetting by modifying learning across classes, a dual distribution memory dynamically manages clean and adversarial samples to imitate realistic data shifts. Despite its novelty, adaptability may be limited in real-time environments due to the complexity of creating adversarial samples. Its generalizability also depends on the quality and representativeness of those samples, which may not fully reflect real-world variability. Furthermore, operational feasibility is challenged by high computational needs, particularly in managing dynamic memory and maintaining performance under strict latency limitations. Although the threat model recognizes evolving attacks, practical application would require achieving a balance between responsiveness, resource constraints, and learning effectiveness.
\begin{limitationbox} In IDS literature, adversarial training is still a common defense strategy.   However, the majority of methods are limited in their capacity to adjust to evolving attack techniques since they rely on pre-defined or static adversarial patterns. Their evaluation is often limited to specific attack types or datasets, which raises questions over their generalizability to unknown threats. Furthermore, the computing needs of generating and retraining on adversarial samples challenge operational feasibility, particularly for real-time systems. These limitations emphasize the necessity of lightweight, adaptive, and generalizable defenses that do not rely solely on prior knowledge of attacks or extensive retraining cycles.
\end{limitationbox}
    
\item \textbf{Structure-based defenses:} This group of defenses aim to improve the intrinsic robustness of ML models used in IDS by modifying their internal mechanisms or decision boundaries. Unlike adversarial training strategies, these approaches do not rely on pre-exposure to adversarial examples but reduce the model's  sensitivity to perturbations through architectural alterations or preprocessing layers. They can be broadly categorized as follows.
\begin{itemize}
\item \textit{Ensemble learning:} This line of defense combines multiple models to diversify decision boundaries and mitigate vulnerability to adversarial exploitation. Zhang et al. \cite{zhang2020tiki} integrate MLP, CNN, and C-LSTM models using a majority-voting approach to increase resilience. While this approach improves defense against decision-based attacks, it is less adaptive to novel attack strategies not seen during training. Its performance under evolving threat conditions and dynamic traffic remains uncertain, limiting its generalizability and operational feasibility in diverse scenarios. Additionally, the scenarios examined might not fully represent the complexity of adversarial behavior in the real world, such as rapidly evolving strategies or highly advanced adaptive attacks. Continuous updating and retraining is required to maintain its effectiveness in operational settings.

Additionally, McCarthy et al. \cite{mccarthy2023defending} develops hierarchical classification to reduce the attack surface exposed to adversaries. The defense has two-layer structure where broad categories are classified first, followed by more detailed decisions. This multi-layered strategy aims to isolate and contain adversarial influence. According to the authors, attackers have only limited understanding of the system in a gray-box threat model.  Although this approach shows promise in restricting attacks, the static hierarchical structures constrain its adaptability because they might not react well to novel or evolving attack tactics. The method's assumptions on feature perturbation may also limit its generalizability in larger or dynamic network settings. Regarding operational feasibility, the complexity of maintaining hierarchical logic and consistent classification across layers presents overhead and potential fragility in real-time situations.

Another study proposed FGMD (Feature Grouping and Multi-model Fusion Detector) \cite{jiang2022fgmd}; an ensemble-based defense technique designed for IoT intrusion detection. Through feature grouping and the incorporation of several classifiers with distinct architectures, each trained on a separate feature subset, the system improves robustness. FGMD aims to identify adversarial samples without relying on the model internals by operating under a black-box threat model with evasion attacks throughout the inference phase.  Although the method respects IoT-specific limitations and exhibits promise in controlled settings, it is unclear how well it will generalize across other attack types and adapt to sophisticated adversarial methods.  Furthermore, the practicality of real-world operations has not been validated, particularly under dynamic traffic and limited resource conditions.

\item \textit{Input transformation:} By preprocessing inputs to eliminate or neutralize perturbations before they reach the model, input transformation techniques aim to defend against adversarial attacks.  These defenses apply transformations like feature squeezing, bit-depth reduction, noise filtering, or input quantization to mitigate the impact of adversarial examples at test time, assuming that they differ from clean inputs in subtle statistical ways. This strategy is demonstrated by one study \cite{roshan2024boosting}, which suggests a two-phase defense that includes Feature Squeezing (FS) for inference and Gaussian Data Augmentation (GDA) for training.  The main defensive element is FS, which is used at test time to compress feature space and reduce vulnerability to adversarial perturbations, even if GDA improves generalization by adding noise-based unpredictability in training data.  Although the defense is effective against white-box attacks like the Carlini \& Wagner (C\&W) approach, its adaptability and generalizability to dynamic, real-world black-box scenarios are limited due to its limited focus on known attack types and reliance on static datasets.
\item \textit{Detection and rejection:} This category either apply mitigation or reject adversarial inputs before they reach the target model. It usually employs auxiliary detectors to detect anomalies in features, predictions, or distributions, aiming to prevent attacks without modifying the core model. The TND framework \cite{hu2025tnd}, which uses a two-stage procedure, is an illustration of this approach. It first detects possible adversarial samples using Locality Sensitive Hashing (LSH), and then it purifies them before classifying them using a contrastive learning-enhanced denoising autoencoder. The system's limitations include the potential to misclassify benign traffic, the need for human threshold modification to manage distribution shifts, and the risk of poor generalization to novel or highly adaptive threats, despite its effectiveness in detecting specific assaults and retaining low computing overhead. Despite these challenges, TND's non-invasiveness makes it a possible option for practical deployment under reasonable restrictions.

Another study introduces FID-GAN \cite{de2020intrusion}, a fog-based, unsupervised IDS optimized for cyber-physical systems (CPSs). It is classified as a detection and rejection defense since it uses reconstruction loss to identify anomalous or adversarial behavior.  The system detects deviations that probably signal attacks by mapping inputs to a latent space and calculating how well they can be reconstructed. An encoder is trained to accelerate reconstruction loss computation to satisfy real-time restrictions, resulting in a detection latency improvement of more than 5.5× over baselines. Although the method works well in low-latency CPS situations, its reliance on reconstruction quality may restrict its generalizability in a wider range of network scenarios and its adaptability when dealing with heterogeneous data sources.

Additionally, Pawlicki et al. \cite{pawlicki2020defending} propose a defense strategy based on detection and rejection that aims to identify adversarial attempts on ML-based IDS without modifying the model architecture. The technique specifically targets evasion attacks by identifying and rejecting unusual inputs. While promising, its scope is limited as it is not able to react to changing hacker behaviors. Also, the paper does not elaborate on its real-world deployment feasibility, raising concerns about scalability and computational overhead in dynamic operational environments.
\item \textit{Feature-space regularization:} This category covers defenses that maintain stability in the feature space, aiming to smooth the model's decision boundaries.    These approaches usually use training-time strategies that generate consistent behavior from the model under minor perturbations, which enhances generalization and minimizes sensitivity to adversarial inputs. For instance, Jmila et al. \cite{jmila2022adversarial} explore Gaussian data augmentation as a lightweight regularization technique. By integrating Gaussian noise into the original samples, the training dataset is expanded in this method, which enables the model to generate stable predictions even with slight input variations.  Although the technique strengthens resistance to some noise-based attacks, its effectiveness is inconsistent: the defense shows dataset dependency, performs differently across classifiers, and can actually experience performance degradation, as observed on UNSW-NB15, demonstrating its limited adaptability and the trade-offs between accuracy and robustness.

Moreover, Chen et al. \cite{chen2023feddef} propose FedDef; an optimization-based input perturbation countermeasure created for Federated Learning (FL)-based Network IDS. It aims to prevent privacy leakage and adversarial evasion by injecting perturbations into traffic features. The objective is to reduce gradient distance (ensuring model utility) while increasing input distance (improving privacy). Although FedDef outperforms several baselines in terms of privacy protection and maintains minimal accuracy loss, its adaptability and generalizability may be limited by hyperparameter sensitivity and task-specific tuning. Furthermore, the optimization's processing complexity might make real-time deployment challenging.
\item \textit{Manifold projection:} The objective of this technique is to identify adversarial samples by analyzing their geometric behavior in feature space. The MANDA \cite{wang2022manda} framework exemplifies this method by analyzing the observation that adversarial perturbations frequently remain close to their original benign inputs despite generating misclassification. In order to differentiate between legitimate traffic and subtly modified adversary inputs, the system projects samples onto a learned manifold. Although MANDA boosts robustness, it is less adaptive if attackers modify their perturbation technique to exploit different geometric properties. Moreover, since the strategy relies on specific assumptions about adversarial behavior, its generalizability is limited. Real-world applicability also presents c hallenges such as adaptability to dynamic network conditions, retraining requirements, and computational costs.
\end{itemize}
\begin{limitationbox} Architecture-based defenses are still largely unexplored in the context of NIDS, despite their conceptual appeal. The majority of research focuses on single-case evaluations using specific models or attack types, frequently ignoring to confirm resilience in a variety of dynamic threat scenarios.
\end{limitationbox}
\end{enumerate}
Despite the variety of adversarial defenses that have been explored, the overall landscape remains relatively limited and uneven, especially for ML-based NIDS. There is a heavy reliance on adversarial training, but it often suffers from poor adaptability to novel attacks, limited generalizability beyond the training distribution, and high computational. Some categories like manifold-based defenses \cite{stutz2019disentangling}, gradient masking techniques \cite{chen2022adversarial}, and input perturbation strategies \cite{ali2022evaluating} are rarely implemented or explicitly studied in this field. Due to the challenge of applying geometric assumptions to structured traffic data, manifold projection has not been widely used in NIDS research, despite its potential for capturing underlying data structures. Gradient masking, often criticized for providing superficial resilience, is often avoided because of its susceptibility to adaptive attacks. Input perturbation techniques, while extensively explored in the image classification context (where small pixel-level noise can maintain semantics), are rarely adapted to NIDS due to the strict semantic and functional constraints inherent in network traffic data. 

\begin{takeawaybox} These limitations highlight a significant challenge: defense mechanisms, particularly those that can generalize across attack types, adapt to evolving threat landscapes, and function effectively in real-world scenarios, are still underdeveloped despite the increased awareness of adversarial threats.
\end{takeawaybox}

%% file: threat_model.tex
\begin{figure*}[htbp]
  \centering
  \includegraphics[trim=1cm 6cm 2cm 1cm, clip, width=0.8\textwidth]{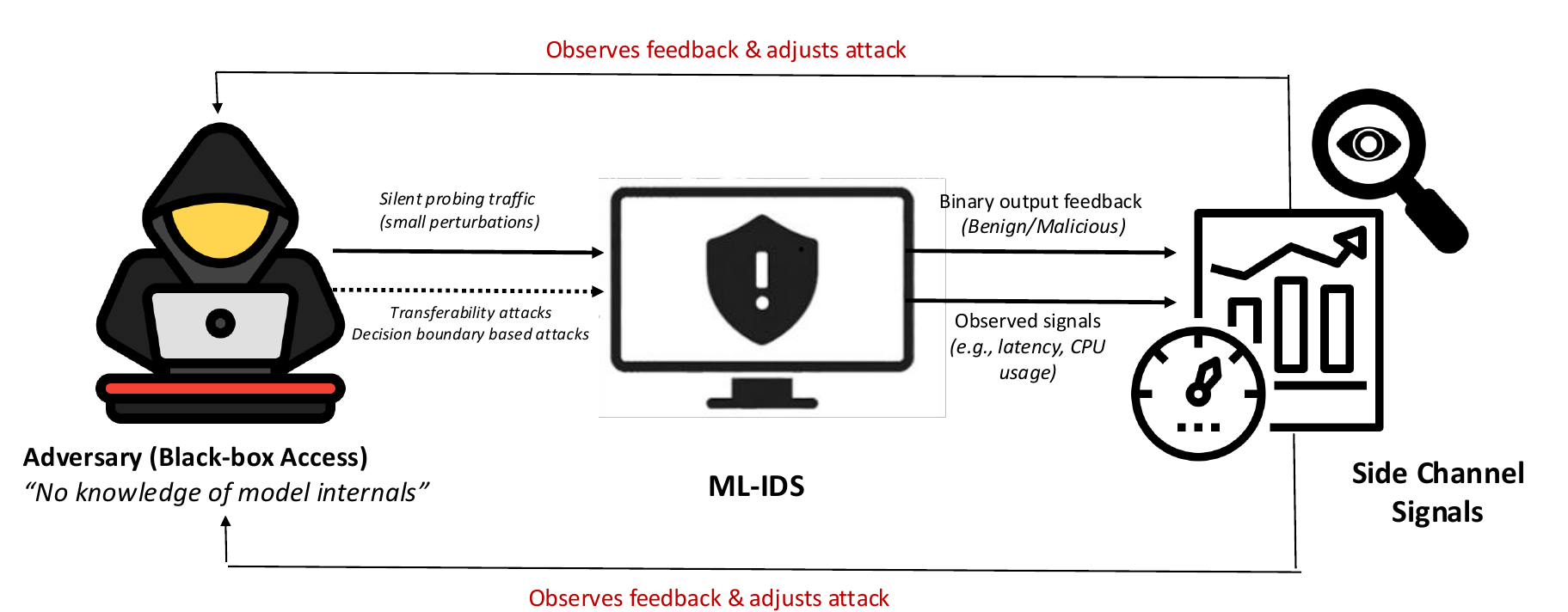}
  \caption{\textbf{Threat Model Overview:} The adversary operates in a black-box setting and interacts with the ML-based IDS by injecting crafted traffic. Instead of accessing internal model outputs, the attacker passively observes indirect system feedback, which is used to infer sensitive decision boundaries through probing for example.}
  \label{fig:threat_model}
\end{figure*}

\section{Threat Model}
\label{sec:threat_model}

This section presents the adversarial setting considered in our study, describing the attacker’s capabilities, limitations, and goals. The threat model is based on realistic assumptions, reflecting how attackers operate in black-box environments without access to internal IDS parameters. The overall threat model is depicted in Figure \ref{fig:threat_model}, where the adversary observes behavioral feedback signals without triggering detection by sending carefully crafted traffic to the system. These indirect observations allow the attacker to probe and learn the behavior of the system while being undetected.

\subsection{Assumptions and Attacker Capabilities
}
We assume that the core model, architecture and training data of the IDS are not accessible to the attacker. He lacks access to gradients, probabilities, or confidence ratings since the model is described as a "black box." However, the attacker is assumed to:

\begin{itemize}
    \item Have the ability to introduce carefully generated network traffic into the monitored environment.
    \item Observe the system's output as a binary decision (malicious vs. benign, for example), or infer feedback from outside indications like system responsiveness, latency, or other non-intrusive behavioral signals.
    \item Operate silently, avoiding large-scale or aggressive behaviors that might trigger traditional alarms.  
\end{itemize}

This scenario represents adversaries who relies on indirect observations to estimate the system’s decision boundaries without triggering detection measures. We define this as a probing phase, during which the attacker iteratively adjusts input patterns to identify features affecting the model’s decision.

\begin{takeawaybox}Both our technique and AFP defense are model-agnostic, in line with the black-box circumstances. Our method is not restricted to any specific model, although for experimental evaluation, we instantiate the target IDS as a Random Forest, which is commonly used in NIDS research.\end{takeawaybox}

\subsection{Attack Scenarios Considered}
Two types of black-box adversarial attack strategies are implemented to evaluate the proposed defense:

\begin{enumerate}
    \item \textbf{Silent Probing Attacks:} These are non-disruptive, passive attacks. The attacker observes the behaviors of the external system while gradually injecting traffic with slight perturbations.   They make an effort to identify which features are essential to the model's predictions through analyzing behavioral patterns over time. Without triggering alarms, the attacker may create a surrogate model or uncover evasion paths using statistical methods (such as change-point detection) and feedback loops.
    \begin{takeawaybox}
        The silent probing methodology used in this work, which was designed to reflect realistic black-box adversarial scenarios, is detailed in \cite{ennaji2026vulnerability}, including feature sensitivity extraction using change-point detection and causal analysis.
    \end{takeawaybox}
     \item \textbf{Transferability-Based Attack:} In this scenario, the adversary approximates the behavior of the target IDS by training a surrogate model with readily available or synthesized data. Adversarial examples are applied against this surrogate and then transferred to the target system, exploiting the tendency of some adversarial manipulations to remain effective across several models \cite{adeke2025investigating}. 
     \item \textbf{Decision Boundary-Based Attack:} By iteratively adjusting input features, the attacker seeks to directly investigate and exploit the target IDS's classification boundary. Identifying minor perturbations that cause a sample to change from one class to another (from harmful to benign) is the goal \cite{finlay2019logbarrier}.
\end{enumerate}

Given the attacker’s ability to infer crucial system behavior based on indirect observations under black-box settings, traditional detection-based defenses are insufficient. This motivates the necessity for a proactive and generalizable approach that functions independently of the classifier’s internal logic. In response, we propose Adaptive Feature Poisoning (AFP) defense, to trick adversarial inference without degrading IDS performance. Section \ref{sec:defense} provides a detailed overview of the AFP's architecture, integration, and implementation.

%% file: defense.tex
\section{Proposed Defense- Adaptive Feature Poisoning (AFP)}\label{sec:defense}

Adaptive Feature Poisoning (AFP) is implemented as an additional layer within the IDS framework that runs in parallel with the main detection approach. While the IDS is examining incoming traffic for potential threats, AFP proactively intervenes to corrupt the feedback loop exploited by adversaries during the reconnaissance phase.

To prevent adversaries from reliably gathering information through side-channel indicators (e.g., processing delays, memory usage, or response time), AFP modifies incoming traffic features before classification, injecting subtle, context-aware noise that deceives the attacker without affecting legitimate detection.

The overall workflow of the proposed AFP is presented in Figure \ref{fig:pf}, which depicts its integration within the IDS pipeline and the attacker's corrupted feedback loop. The detailed operational logic is also described by an algorithm, which outlines the adaptive profiling, perturbation, and escalation strategy implemented by AFP.

AFP differs from other defenses in its reliance on a profiling and analysis pipeline that dynamically adapts perturbation strength and target features. This process consists of the following steps:

\begin{enumerate}
\item \textbf{Monitoring and Detection:}

In the first phase, AFP continuously monitors input traffic and system-level side-channel indications. By gathering a statistical window of benign feature values and associated side-channel metrics during an initialization phase, AFP creates a baseline profile $\mathcal{B}_{ref}$ of normal traffic behavior to serve as a reference point.

\begin{figure*}[htbp]
  \centering
  \includegraphics[trim=5.1cm 5cm 5cm 6cm, clip, width=0.8\textwidth]{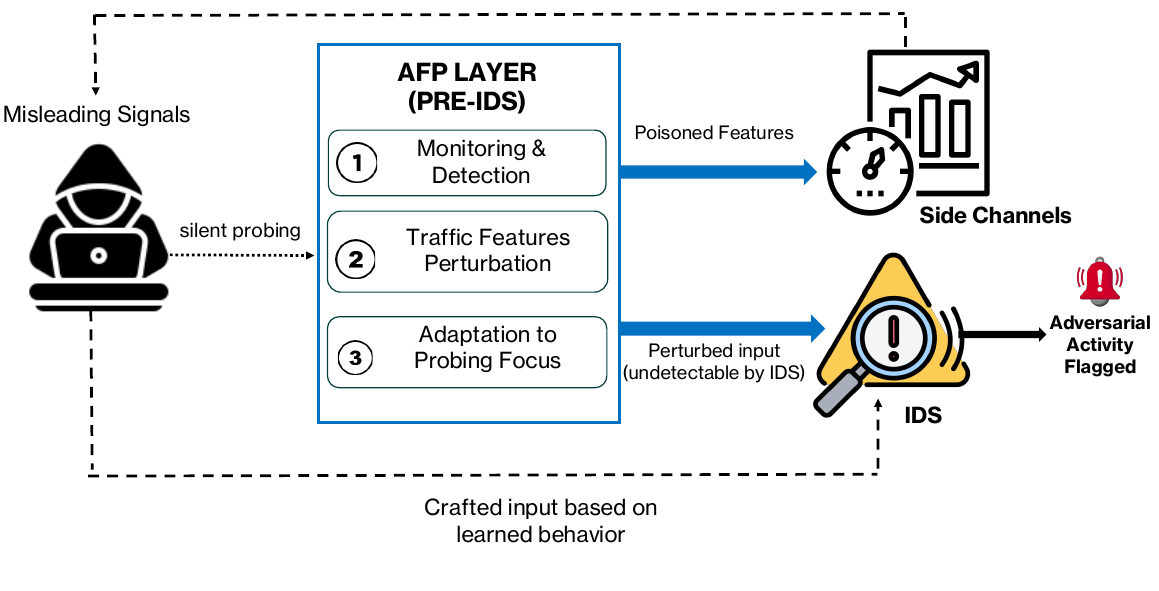}
  \caption{Overview of the Adaptive Feature Poisoning (AFP) defense framework.}
  \label{fig:pf}
\end{figure*}

This baseline captures the distribution of selected traffic features (e.g., Duration, BytesPerSec, and PktsPerSec) and related side-channel signals, such as CPU usage and response time.

Let:
\begin{itemize}
\item $\mathcal{X}{ref} = { X{t-n}, \dots, X_{t} }$ be the baseline traffic feature window
\item $\mathcal{S}{ref} = { S{t-n}, \dots, S_{t} }$ be the baseline side-channel observations
\end{itemize}
While operating in real time, AFP continuously collects:
\begin{itemize}
\item A sliding window $\mathcal{X}_{obs}$ of the most recent traffic features
\item A parallel window $\mathcal{S}_{obs}$ of the most recent side-channel signals
\end{itemize}
AFP applies a change-point detection algorithm \textit{(Binary Segmentation \cite{aminikhanghahi2017survey})} over $\mathcal{S}{obs}$ to identify sudden shifts in side-channel behavior. A change point $t^*$ is detected when the distribution of $\mathcal{S}{obs}$ deviates significantly from $\mathcal{S}_{ref}$ \textit{(e.g., by means of mean / variance/ variance break or cumulative sum analysis)}. 

Hence, this suggests potential manipulation of the system or probing activity.

\begin{algorithm}[t!]
\renewcommand{\thealgocf}{} 
\caption{\textbf{Adaptive Feature Poisoning (AFP)}}

\SetAlgoNoLine
\SetAlgoNlRelativeSize{0}
\SetNlSty{}{}{}
\DontPrintSemicolon
\label{alg:afp}
\KwIn{
    Incoming traffic features $X$,\\
    Side-channel signal $S$ (e.g., response time),\\
    Feature set $\mathcal{F}$,\\
    Baseline profile $\mathcal{X}_{\text{ref}}$,\\
    Threshold $\theta$,\\
    Base perturbation $\epsilon_{\text{base}}$,\\
    Scaling factor $\alpha$
}
\KwOut{
    Perturbed traffic $X'$
}

\textbf{Step 1: Monitoring and Detection} \\
Compute current sliding window $\mathcal{X}_{\text{obs}}, \mathcal{S}_{\text{obs}}$ \;
Apply change-point detection on $\mathcal{S}_{\text{obs}}$ \;
\eIf{change detected ($\mu_S > \theta$)}{
    Compare $\mathcal{X}_{\text{obs}}$ to $\mathcal{X}_{\text{ref}}$ \;
    Identify $\mathcal{F}_{\text{deviated}} \subseteq \mathcal{F}$ based on statistical deviation \;
}{
    $\mathcal{F}_{\text{deviated}} \leftarrow \emptyset$ \;
}

\textbf{Step 2: Perturbation Initialization} \\
Initialize $X' \leftarrow X$ \;

\textbf{Step 3: Adaptation to Probing Focus} \\
\ForEach{$f_i \in \mathcal{F}$}{
    \eIf{$f_i \in \mathcal{F}_{\text{deviated}}$}{
        Compute deviation score $\delta_i$ \;
        Compute adaptive perturbation: \\
        \Indp $\epsilon_i = \epsilon_{\text{base}} + \alpha \cdot \delta_i$ \;
        Apply perturbation: \\
        $X'_{f_i} = X_{f_i} + \mathcal{U}(-\epsilon_i, \epsilon_i)$ \;
        \Indm
    }{
        Apply low-level baseline perturbation: \\
        $X'_{f_i} = X_{f_i} + \mathcal{U}(-\epsilon_{\text{base}}, \epsilon_{\text{base}})$ \;
    }
}
\Return $X'$
\end{algorithm}

When AFP identifies a change, it determines which features present the most statistical deviation by comparing the baseline $\mathcal{X}{\text{ref}}$ with the observed traffic window $\mathcal{X}{\text{obs}}$.  Lightweight comparison measures, like mean or variance shifts, are used for this, enabling the system to identify features that the adversary is probably targeting.

\begin{takeawaybox} These features are selected as under attacker focus and transferred to the following step for targeted perturbation. \end{takeawaybox}

\item \textbf{Traffic Features Perturbation:} 

AFP starts its perturbation method after detecting a behavioral deviation in system-side indicators and identifying the traffic features that diverge from their expected baseline. The purpose is to silently disrupt the attacker's capabilities to collect reliable feedback, while maintaining the integrity of the IDS's operation.

AFP injects small, randomized perturbations into the subset of features previously identified as showing abnormal behavior. These features are likely the ones the attacker is targeting to estimate decision boundaries or threshold behavior in a black-box setting.

The degree of deviation of each feature from its baseline is used to adaptively determine the perturbation strength. Let $f_i$ be a selected feature and $\epsilon_i$ its corresponding perturbation strength, then:

\[
X'_{f_i} = X_{f_i} + \mathcal{U}(-\epsilon_i, \epsilon_i)
\]

Where: 

\begin{itemize}
\item $X_{f_i}$: the original value of feature $f_i$, 
\item $\epsilon_i$: the adaptive perturbation strength for $f_i$, 
\item $\mathcal{U}(-\epsilon_i, \epsilon_i)$: a value sampled from a uniform distribution within the range $[-\epsilon_i, \epsilon_i]$,
\item $X'_{f_i}$: the perturbed value forwarded to the IDS for classification
\end{itemize}

The deviation score of the feature, which indicates how unusual its recent values are in relation to the reference profile, is scaled proportionately to the magnitude $\epsilon_i$. This enables AFP to apply more aggressive noise to features under targeted probing, while leaving others lightly manipulated or unperturbed.

These perturbations are carefully bounded to ensure they do not degrade the system’s classification performance. For instance, a legitimate packet size of 600 bytes might be reported as 603 bytes; this alteration preserves semantic accuracy but poisons the attacker's statistical inference.

\begin{takeawaybox} AFP functions continually, injecting low-level perturbations throughout the observed features as a proactive deterrent even in the absence of explicit change points.  Following confirmation of a deviation, perturbation becomes adaptive, more aggressively focusing on the attacker's priority.
\end{takeawaybox}

\item \textbf{Adaptation to Probing Focus:}

In addition to continuous low-intensity perturbation, AFP applies an adaptive escalation strategy that more precisely targets the attacker’s focus. Once a deviation is confirmed via change-point detection and feature-level deviations are measured, AFP prioritizes features showing repeated or focused abnormal patterns across different traffic windows.

For every feature $f_i$, AFP dynamically modifies the perturbation strength $\epsilon_i$ according to the consistency and frequency of its deviation. Let $\delta_i$ be the feature $f_i$'s cumulative deviation score over recent windows. The following is the corresponding increase in perturbation strength caused by AFP:

\[
\epsilon_i = \epsilon_{\text{base}} + \alpha \cdot \delta_i
\]

Where:
\begin{itemize}
\item $\epsilon_{\text{base}}$: the minimum perturbation level applied to all features,
\item $\delta_i$: the deviation magnitude for feature $f_i$ compared to its baseline profile,
\item $\alpha$: a scaling factor that identifies how aggressively AFP reacts to persistent probing on $f_i$
\end{itemize}

This adaptive tactic ensures that the more an attacker attempts to probe a specific feature (e.g., Duration, BytesPerSec), the less reliable the resulting feedback becomes. The attacker may observe delayed or noisy responses that are not precisely linked to any internal decision boundary, effectively disrupting surrogate model building or decision threshold estimation.

\begin{takeawaybox} All perturbations remain bounded and are applied only to non-critical dimensions that do not directly affect classification decisions. This design guarantees that AFP will continue to be lightweight, invisible, and appropriate for real-time deployment.
\end{takeawaybox}

\end{enumerate}

%% file: discussion.tex
\section{Results and Discussion}
\label{sec:discussion}
A thorough assessment of the proposed AFP defense in a ML-based IDS is provided in this section, focusing on how well it defends against a variety of realistic black-box adversarial attacks. It first describes the experimental setting. The following subsections evaluate the effectiveness of selectively-triggered AFP, its robustness and generalizability to various attack types, and assess its computational overhead to show how lightweight it is. Lastly, we contextualize the advantages and practical contributions of AFP by providing a comparison with well-known adversarial defenses. This multifaceted evaluation presents new insights into the real-world deployability and security trade-offs of adaptive adversarial countermeasures in existing IDS.

\subsection{Experimental Setup}
Our experimental evaluation was conducted using the CSE-CIC-IDS2018 dataset, a recent and realistic benchmark that includes modern network traffic and a variety of sophisticated attack scenarios. 

\begin{itemize}
\item \textbf{Dataset:} More than 16 million network flows covering a broad range of modern benign and attack classes (such as DDoS, brute-force, online attacks, and infiltration) were collected over five days for the CSE-CIC-IDS2018 dataset. Each flow is labeled as benign or malicious, making it particularly suited for evaluating IDS robustness in practical, up-to-date environments \cite{ennaji2026vulnerability}.

\item \textbf{Preprocessing and Feature Selection:} Missing value records were excluded. Unit variance and zero mean have been employed to standardize continuous features. In order to maximize IDS performance and computational efficiency, a subset of relevant features was selected based on feature importance analysis and domain knowledge.

\item \textbf{Target IDS System:} TThe IDS was implemented as a Random Forest (RF) classifier, a well-established model in the field. Model hyperparameters were tuned via grid search for optimal performance. Model hyperparameters were tuned via grid search for optimal performance.

\item \textbf{Training and Validation:} To prevent bias, the dataset was divided into distinct training and testing sets while preserving class balance. F1 score, recall, accuracy, and precision were used to evaluate performance.

\item \textbf{Adversarial Attack Generation:} For each attack scenario, adversarial samples were generated as detailed in Section~\ref{sec:threat_model}. Silent probing attacks mimic highly skilled adversaries who use subtle and undetected queries to progressively infer system boundaries.
Transferability-based attacks represent the practical threat that adversaries can create effective evasive samples using surrogate models trained on comparable but different data, which is an increasingly relevant challenge in real deployments. Decision boundary-based attacks target weaknesses even in cases where attackers have indirect or limited access to features by directly exploiting areas of classifier ambiguity. These were used to assess IDS robustness with and without AFP strategy.
\end{itemize}

\begin{takeawaybox}
The findings of this experiment are ensured to be operationally realistic and demonstrate how well AFP defends against current, real-world adversarial threats.
\end{takeawaybox}

\subsection{Effectiveness of Selectively-Triggered AFP} 
In all three attack situations, AFP was used as intended, being selectively activated based on change-point detection and side-channel monitoring. In real world, this mechanism only triggered for 100 of the 1,314,720 network flows (less than 0.01\% of all traffic), ensuring that the vast majority of samples were processed without perturbations.  This adaptive, targeted deployment offers strong adversary resilience with slight impact on legitimate traffic, making it a scalable and useful protection strategy for practical IDS deployments.

\begin{figure}[h]
    \centering
    \includegraphics[width=0.9\linewidth]{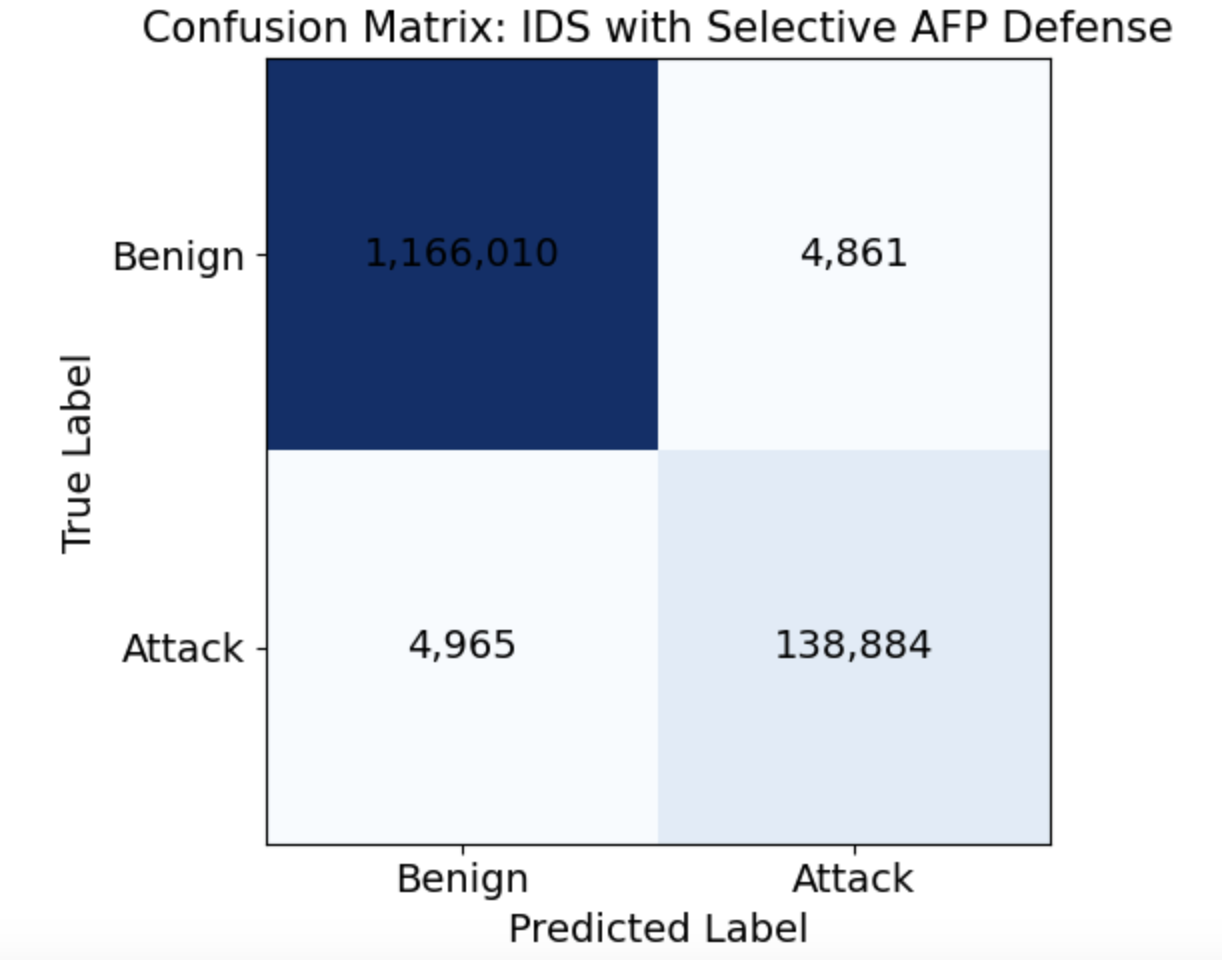}
    \caption{Confusion matrix for the IDS with selectively-triggered AFP approach.}
    \label{fig:conf}
\end{figure}

Performance under selective AFP was excellent, as illustrated by the confusion matrix in figure \ref{fig:conf} and summarized in the corresponding table \ref{tab:afp-results}: 

\begin{itemize}
\item Overall accuracy achieved 99.3\% after adversarial attacks.
\item For benign samples, recall and precision were 1.00.
\item For attack samples, recall and precision were higher than 97\%.
\item In comparison to the size of the dataset, only 4,965 attack samples out of 143,849 were overlooked (false negatives), and 4,861 benign samples were incorrectly identified (false positives).
\end{itemize}

This high-fidelity result shows that AFP maintains nearly all of the IDS's detection capabilities against both benign and attack traffic, while providing strong resilience to adversarial manipulation, when activated only in suspicious or unusual scenarios.  
In contrast to more aggressive (always-on) AFP settings, the selective activation approach avoids the danger of compressing all attack points into the benign class. Therefore, the IDS remains highly reliable and effective in realistic threat environments, with negligible performance overhead.

{
\scriptsize                         
\setlength{\tabcolsep}{2.5pt}       
\renewcommand{\arraystretch}{0.8}   

\begin{table*}
\centering
\caption{Performance of Selectively-Triggered AFP Defense}
\label{tab:afp-results}

\begin{tabular}{lcccc}
\toprule
\textbf{Scenario} & \textbf{Accuracy} & \textbf{Benign Recall} & \textbf{Attack Recall} & \textbf{AFP Coverage} \\
\midrule
Triggered AFP (ours) & 0.9925 & 1.00 & 0.97 & 0.01\% \\
\bottomrule
\end{tabular}
\end{table*}

}

\begin{takeawaybox}These findings demonstrate the significance of evidence based, adaptive triggering for adversarial defenses in operational ML-based intrusion detection systems.
\end{takeawaybox}

\subsection{Generalizability and Robustness Across Attack Types}

\textbf{Attack Baseline Results:}

With an overall accuracy of 99.25\% and nearly perfect detection rates for both benign and attack traffic (benign recall: 1.00, attack recall: 0.97), the Random Forest-based intrusion detection system demonstrated strong baseline accuracy before adversarial manipulation.  Under typical, non-adversarial circumstances, this performance shows how effective the model is.

\begin{figure*}[htbp]
    \centering
    \includegraphics[width=0.8\linewidth]{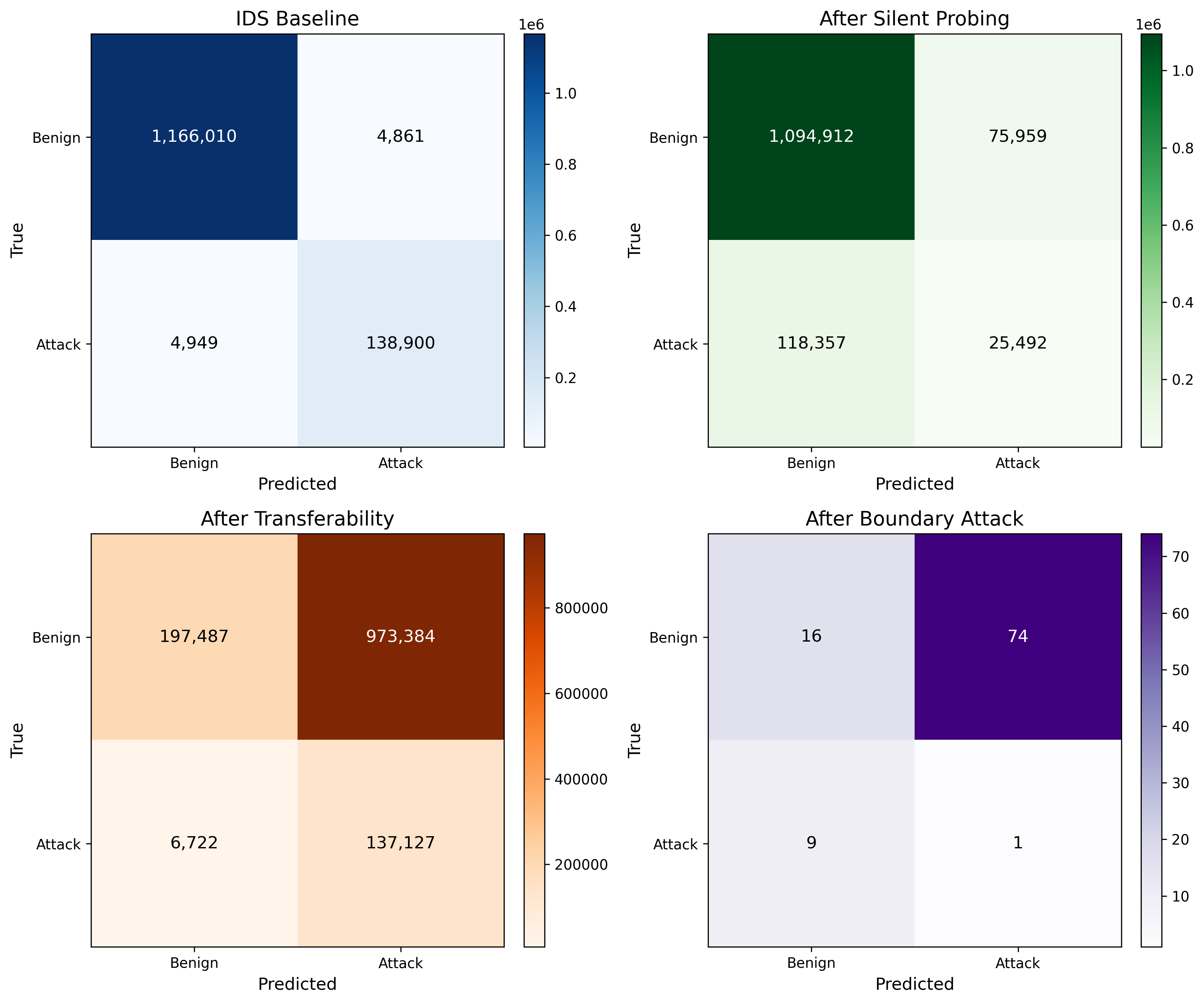}
    \caption{
        Confusion matrices for the IDS under (a) baseline conditions, (b) silent probing attack, (c) transferability-based attack, and (d) decision boundary attack. Each subplot shows the number of correctly and incorrectly classified benign and attack samples, highlighting the degradation in IDS detection performance under adversarial scenarios.
    }
    \label{fig:attack-baseline-matrix}
\end{figure*}

However, all three black-box attack scenarios severely limited detection, as seen in Table~\ref{tab:attack-results}.  With more than 75,000 benign samples incorrectly identified as attacks and more than 118,000 attacks ignored, the silent probing attack decreased accuracy to 85.2\% and attack recall to just 18\%.  Even worse was the transferability-based attack, which reduced accuracy to 25.5\% and almost reversed recall, misclassifying almost all of benign samples as attacks while the attack class had a recall of 0.95.  With a high evasion success rate. For the decision boundary attack, it has dropped the accuracy to 17\%, misclassifying the great majority of both classes.

\begin{table*}[ht]
\centering
\caption{IDS performance under attack (Before AFP).}
\label{tab:attack-results}
\begin{tabular}{lcccc}
\hline
\textbf{Attack Type }& \textbf{Accuracy} & \textbf{Benign Recall} & \textbf{Attack Recall} & \textbf{Support} \\
\hline
Silent Probing      & 0.852  & 0.94  & 0.18 & 1,314,720 \\
Transferability     & 0.255  & 0.17  & 0.95 & 1,314,720 \\
Boundary Attack     & 0.170  & 0.18  & 0.10 & 100 \\
\hline
\end{tabular}
\end{table*}

These findings, supported by the detailed confusion matrices illustrated in Figure~\ref{fig:attack-baseline-matrix}, highlight the practical feasibility and severity of black-box adversarial attacks against machine learning-based IDS.

\textbf{AFP Defense Results:}

\begin{figure*}[ht]
    \centering
    \includegraphics[width=0.8\textwidth]{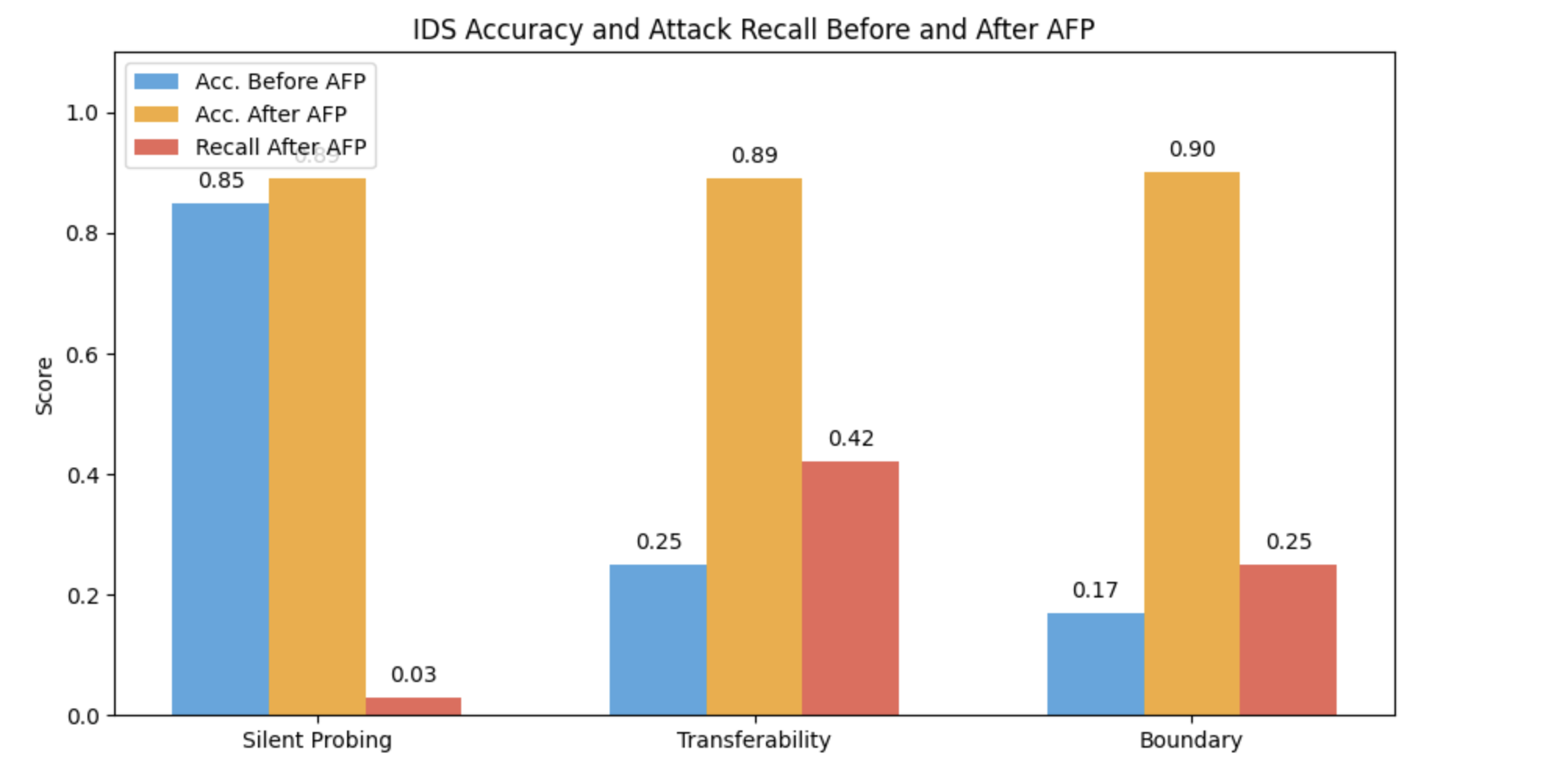}
    \caption{IDS accuracy and attack recall before and after AFP.
    }
    \label{fig:after}
\end{figure*}

Upon activation of AFP approach, the IDS robustness demonstrates notable increases against all considered adversarial methods, as it is depicted in Table~\ref{tab:afp-defense-results} and Figure \ref{fig:after}. The following analysis explores the mechanisms driving these findings and examine their relevance for each attack vector.

\begin{table*}[ht]
    \centering
    \caption{IDS performance before and after AFP for each attack scenario.}
    \label{tab:afp-defense-results}
    \begin{tabular}{lccc}
        \hline
        \textbf{Attack} & \textbf{Acc. Before} & \textbf{Acc. After AFP} & \textbf{Attack Recall After AFP} \\
        \hline
        Silent Probing     & 0.8522 & 0.8906 & 0.03 \\
        Transferability    & 0.2545 & 0.6154 & 0.42 \\
        Boundary           & 0.1700 & 0.9000 & 0.01 \\
        \hline
    \end{tabular}
\end{table*}

\begin{itemize}
\item \textbf{\textit{Silent Probing Attack:}}

Silent probing is considered as a realistic black-box adversarial threat in which attackers slightly perturb network traffic to map the IDS decision boundary while avoiding detection. Without defense, this technique drops the IDS accuracy as attack samples imitate legitimate traffic, allowing for significant evasion. 

When AFP is integrated, the defensive environment changes substantially.   AFP directly disrupts the attacker's learning process by adaptively introducing small, context-aware perturbations into a selection of sensitive features only when probing is detected. This breaks the adversary’s feedback loop: the feedbacks the attacker observes are no longer consistent with the true decision boundary, which frustrates boundary inference and diminishes attack adaptation. 

AFP increases accuracy quantitatively from 85.2\% to 89.1\%, which is a significant improvement given the high-stakes nature of large-scale network defense.  Furthermore, this enhancement is made without randomly disrupting all traffic; instead, the selective, triggered nature of AFP ensures that normal operations remain largely unaffected, maintaining the reliability of the system.

The subtle impact on attack recall (0.03 after AFP) must also be recognized. Instead of reducing all attacks to benign, which would hide continuous adversarial activity, AFP allows some more severe or reckless adversarial attempts to remain detectable. This property is operationally valuable; as attackers become overconfident or escalate their probing due to misleading feedback, they reveal themselves to detection, providing defenders with actionable signals for incident response and escalation monitoring.

\item \textbf{\textit{Transferability-based Attack:}}

When AFP is activated, the dynamic shifts. AFP adaptively perturbs traffic in response to detected adversarial patterns, specifically targeting sensitive features in a way that interferes with the transferability of crafted adversarial samples. Consequently, the accuracy of the IDS increases sharply from 25.4\% to 89.1\%, regaining a significant portion of its initial discriminatory  power. Transferability-based attacks are generally difficult to defend against without the employment of sophisticated or computationally costly countermeasures, which makes this recovery notable. 

The critical insight is that AFP disrupts the implicit "agreement" between the distributions of the surrogate and target models: even small, context-aware disturbances weaken the effectiveness of transferred adversarial samples, making them unable to perform evasion.  In practical terms, this means that attackers cannot consistently use surrogate-based tactics without setting off defenses and running a risk of being detected.

Moreover, a realistic trade-off can be observed in the moderate recall value of 42\% after AFP, which provides strong protection with few false positives by selectively perturbing features only when probing is detected. In the real world, this means that the IDS maintains intercepting a significant portion of sophisticated attacks while boosting the effort and exposure risk of attackers—all without adversely affecting benign traffic.

\item \textbf{\textit{Decision Boundary-based Attack:}} 

Decision boundary attacks pose a challenging risk for black-box IDS deployments because they exploit the classifier's margins by creating adversarial samples that specifically search for points where the model's predictions deviate.   This approach proved to be highly effective in our baseline study, lowering IDS accuracy to just 17\% while successfully masking the majority of attack samples as benign.

Similarly to the previous attacks, AFP's adaptive perturbations introduce controlled unpredictability which hinders evasion by interfering with the attacker's search for the true boundary.
When compared to almost zero without defense, the improvement in attack recall demonstrates that AFP significantly restores the IDS's detection capability against boundary-based threats. Operational dependability is maintained by selective activation, which implies that benign traffic is mainly unaffected. This trade-off illustrates AFP's effectiveness further because it forces the attacker to engage in riskier, more detectable activity while preserving strong regular IDS performance.
\end{itemize}

\begin{takeawaybox} Instead of merely blocking attacks at the cost of usability, AFP makes the adversary's environment strategically confusing. By selectively altering features in response to boundary, transferability-based, and probing attacks, AFP deceives attackers and pushes them to take increasingly conspicuous behaviors. This adaptive resilience maintains IDS performance on benign traffic while increasing attacker risk and facilitating more robust, layered detection. This is how AFP transforms traditional adversarial blind spots into observable, manageable threats—precisely the defensive posture required in modern, high-stakes environments.    
\end{takeawaybox}

\subsection{Computational Overhead and Lightweightness}

The highly low computational overhead of the suggested AFP defense is a critical advantage, as shown in presented in Figure \ref{fig:afp_overhead}. In each attack scenario, AFP's additional runtime was minimal.   Only 5.95\% of the overhead was associated with silent probing attacks (4.44s baseline vs. 4.70s with AFP). Even lower was the overhead for attacks based on transferability, which was 2.72\% (9.29s vs. 9.54s). For boundary-based attacks, the overhead was negligible and in some cases slightly negative (-0.77\%) due to timing variability in ultra-fast inference, which we attribute to measurement noise. 

This efficiency results from AFP's selective and on-demand design, which only triggers AFP for a very tiny portion of suspiciously flagged network flows rather than disrupting all of them. in our evaluation, less than 0.01\% of network flows in the large-scale scenarios needed AFP intervention. Consequently, a significant amount of traffic is handled at maximum speed without incurring any extra computational costs.   In comparison to the core IDS inference, only a small number of samples experience light, local feature perturbations, which are computationally insignificant. This targeted approach makes AFP ideally suited for real-time deployment in operational IDS scenarios where resource and latency limitations are critical.

\begin{figure*}[ht]
    \centering
    \includegraphics[width=0.8\linewidth]{figures/afpoverhead.png}
    \caption{Comparison of baseline IDS prediction time and additional overhead introduced by AFP across three attack scenarios. AFP incurs minimal or negligible computational cost, confirming its suitability for real-time deployments.}
    \label{fig:afp_overhead}
\end{figure*}

\begin{takeawaybox}Real-world IDS deployments can incorporate AFP with almost no impact on system responsiveness or prediction speed.\end{takeawaybox}

\subsection{Comparative Analysis with Existing Defenses}

To contextualize the importance of our AFP defense, we compare its design and performance against a range of established adversarial defense approaches, as extensively discussed in Section \ref{sec:review} and summarized in Table \ref{tab:defense_summary}. Each defense presents important trade-offs in terms of adaptivity, operational overhead, and their ability to deaò with unknown/adaptive threats. 

As outlined in Table \ref{tab:defense-comparison}, the following overview highlights AFP’s unique position and practical advantages among existing defenses: 
\begin{itemize}

\item \textbf{Selective, On-Demand Activation:} In contrast to "always-on" techniques, AFP is only activated in reaction to change-point detections or side-channel signs, maintaining normal accuracy and reducing operational overhead.

\item \textbf{Model-Agnostic Integration:} For additional protection, AFP can be implemented in conjunction with already-existing adversarial training or ensemble defenses. It functions as a plug-in layer and does not require model retraining or architectural changes.

\item \textbf{Dynamic Adversarial Disruption:} AFP misleads the attacker, interferes with iterative probing, and increases the cost of successful evasion by adaptively perturbing features only when adversarial behavior is detected.

\item \textbf{Low False Positive Rate:} AFP overcomes one of the key limitations of aggressive or static defenses by maintaining IDS effectiveness and reliability due to the restricted modification of benign samples.
\end{itemize}

\begin{table*}[t]
\centering
\caption{Comparison of AFP with Established Adversarial Defenses}
\label{tab:defense-comparison}

\scriptsize                       
\setlength{\tabcolsep}{3pt}       
\renewcommand{\arraystretch}{0.9} 

\begin{tabular}{%
    p{3.4cm}  
    p{1.8cm}  
    p{1.7cm}  
    p{3.2cm}  
    p{2.6cm}  
    p{2.6cm}  
}
\toprule
\textbf{Defense Approach} &
\textbf{Adaptivity} &
\textbf{Overhead} &
\textbf{Robustness to Unknown Attacks} &
\textbf{Impact on Benign} &
\textbf{AFP Compatibility} \\
\midrule

Adversarial Training \cite{debicha2023adv} &
Low &
High &
Limited &
Moderate/High &
Can be combined \\

Ensemble Learning \cite{mccarthy2023defending} &
Low/Moderate &
Moderate &
Moderate &
Low/Moderate &
Can be combined \\

Input Transformation \cite{roshan2024boosting} &
None &
Mod/High &
Low/Moderate &
Moderate/High &
Can be combined \\

Detection \& Rejection \cite{hu2025tnd} &
Low &
Moderate &
Low &
High &
Can be combined \\

Feature-Space Regularization \cite{chen2023feddef} &
None &
Moderate &
Low/Moderate &
Moderate &
Can be combined \\

Manifold Projection \cite{wang2022manda} &
None &
High &
Low/Moderate &
High &
Can be combined \\

\textbf{AFP (Ours)} &
High (Selective) &
Low &
High &
Very Low &
Works with all \\
\bottomrule
\end{tabular}
\end{table*}

\begin{takeawaybox} AFP stands out by its lightweight and adaptive nature, providing a complementary layer that can operate alongside existing defenses. By activating only when required, our approach preserves system performance, while its dynamic adaptation fills significant gaps left by existing adversarial countermeasures.
\end{takeawaybox}

%% file: conclusion.tex
\section{Conclusion}\label{sec:conclusions}
In order to improve the resilience of machine learning-based intrusion detection systems (IDS) in realistic black-box threat scenarios, this study introduces Adaptive Feature Perturbation (AFP), a novel and lightweight defense mechanism.  We show that AFP can selectively and dynamically disrupt adversarial adaptation, maintain IDS accuracy, and improve the cost and detectability of evasion attempts through extensive experiments on modern adversarial attack strategies, such as transferability-based, decision boundary-based, and silent probing attacks.

Even though AFP provides substantial improvements in adversarial disruption and resilience, there are still some areas that might be addressed.   As future work , we aim to expand AFP into a completely self-learning, self-tuning layer that can autonomously profile evolving network conditions and attacker strategies. Furthermore, we seek to investigate the synergistic integration of AFP with other adaptive defense mechanisms and evaluate its efficacy across a wider range of IDS models and deployment situations, such as online and distributed environments.

By advancing toward self-adaptive, intelligent defense layers, we believe AFP and related techniques can contribute to redefining adversarial resilience in next-generation network security systems.